\documentclass[twocolumn]{aastex62}

\usepackage{graphicx}
\usepackage{amsmath}
\usepackage{graphicx}
\usepackage{amssymb}
\usepackage{color}
\def\modify#1{#1}

\def\hunit{\,\mathrm{km\,s^{-1}Mpc^{-1}}}

\begin{document}

\title{Reaffirming the Cosmic Acceleration without Supernova and CMB}

\author{Xiaolin Luo}
\affiliation{School of Physics and Astronomy, Sun Yat-Sen University, 2 Daxue Road, Tangjia, Zhuhai, 519082, P.R.China}
\author{Zhiqi Huang}
\affiliation{School of Physics and Astronomy, Sun Yat-Sen University, 2 Daxue Road, Tangjia, Zhuhai, 519082, P.R.China}
\email{huangzhq25@mail.sysu.edu.cn}
\author{Qiyue Qian}
\affiliation{School of Physics and Astronomy, Sun Yat-Sen University, 2 Daxue Road, Tangjia, Zhuhai, 519082, P.R.China}
\author{Lu Huang}
\affiliation{School of Physics and Astronomy, Sun Yat-Sen University, 2 Daxue Road, Tangjia, Zhuhai, 519082, P.R.China}

\correspondingauthor{Zhiqi Huang}

\def\page{p_{\rm age}}

\begin{abstract}
  Recent discussions about supernova magnitude evolution have raised doubts about the robustness of the late-universe acceleration. In a previous letter, \citet{Huang:2020mub} did a null test of the cosmic acceleration by using a Parameterization based on the cosmic Age (PAge), which covers a broad class of cosmological models including the standard $\Lambda$ cold dark matter model and its many extensions. In this work, we continue to explore the cosmic expansion history with the PAge approximation. Using baryon acoustic oscillations ({\it without} a CMB prior on the acoustic scale), gravitational strong lens time delay, and passively evolving early galaxies as cosmic chronometers, we obtain $\gtrsim 4\sigma$ detections of cosmic acceleration for both flat and nonflat PAge universes. In the nonflat case, we find a novel $\gtrsim 3\sigma$ tension between the spatial curvatures derived from baryon acoustic oscillations and strong lens time delay. Implications and possible systematics are discussed.
  
\end{abstract}

\section{Introduction}\label{intro}

It has long been known that most of the matter in the universe is dark and, according to the standard interpretation of primordial nucleosynthesis and cosmic microwave background (CMB), non-baryonic. The cold dark matter (CDM) universe has been the standard cosmological scenario until the end of the last century, when the extra dimming of distant Type Ia supernovae was discovered~\citep{Perlmutter:1998hx,Perlmutter:1998np,Schmidt:1998ys,Riess:1998cb}. The standard explanation of the supernova Hubble diagram is that a cosmological constant $\Lambda$, or more generally a dark energy component with negative pressure drives the accelerated expansion of the universe. Since then, the $\Lambda$ cold dark matter ($\Lambda$CDM) model has been confronted with, and passed a handful of observational tests, such as the clustering of galaxies~\citep{Alam:2016hwk}, the weak gravitational lensing of galaxies~\citep{DES1yr_WL}, the updated Type Ia supernova catalogs~\citep{Pantheon, Macaulay_2019}, and most importantly, the full sky CMB temperature and polarization maps measured by the WMAP satellite~\citep{WMAP9} and the Planck satellite~\citep{Aghanim:2018eyx}. 

The remarkable success of the $\Lambda$CDM model, however, is recently challenged by a tension between the local distance-ladder measurement of the Hubble constant, $H_0=74.03\pm 1.42\hunit$~\citep{Riess19}, and the value $H_0=67.4\pm 0.5\hunit$ inferred from CMB + $\Lambda$CDM fitting~\citep{Aghanim:2018eyx}. An independent measurement of $H_0$ from strong gravitational lens time delay prefers a higher $H_0$ value, too, and raises the $H_0$ tension to $5.3\sigma$~\citep{Wong:2019kwg}. While seemingly significant enough to rule out $\Lambda$CDM, the tension in $H_0 \equiv 100 h\hunit$  may also subject to some unaccounted systematics~\citep{EG14, EG20, Handley20}. The debate thus goes on and becomes one of the most topical subjects in recent years.

A less prominent problem of $\Lambda$CDM is the up to $\sim 3\sigma$ tension between the CMB and galaxy weak gravitational lensing constraints on $S_8\equiv \sigma_8\left(\Omega_m/0.3\right)^{0.5}$, where $\sigma_8$ is the root mean square matter density fluctuations within a tophat sphere with radius $8h^{-1}\mathrm{Mpc}$, and $\Omega_m$ is the matter density parameter. This discrepancy is often loosely referred to as the $\sigma_8$ tension, of which the most recent update can be found in the latest weak lensing data release from the Kilo-Degree Survey (KiDS-1000)~\citep{KiDS20}. 

Another potential crisis may stem from the careful scrutiny of the key assumption of supernova cosmology, that the empirically standardized type Ia supernova magnitude does not evolve with redshift. Recently, \citet{Kang:2019azh} suggests that the environmental dependences of supernova magnitude, found in previous works~\citep{Hicken:2009dk,Sullivan:2010mg,Rigault:2013gux,Rigault:2014kaa,Rigault:2018ffm,Roman18, Kim:2019npy}, can all be interpreted as a progenitor-age modulation, which may explain the extra dimming of the distant supernovae without cosmic acceleration. \citet{Rose:2020shp} argued that \citet{Kang:2019azh} had incorrectly extrapolated the statistics on the local early-type host galaxies to host galaxies of all morphological types, whose population age is not significantly correlated with the standardized supernova magnitude. As a quick response \citet{Lee_prepare} pointed out that the statistical analysis by \citet{Rose:2020shp} is affected by regression dilution bias, after correction of which they still find a significant host-age modulation of supernova magnitude. See also \citet{Uddin20} and \citet{Ponder20} for some recent independent voices on this topic.

The on-going debate of Hubble tension, $\sigma_8$ tension, and supernova magnitude evolution is of great importance for the inference of cosmological parameters. Doubts may be raised about the $\Lambda$CDM model and even the late-time acceleration of the universe. In any case, however, as pointed out by ~\citet{Huang:2020mub}, cosmology is unlikely to roll back to the CDM model, because a CDM universe is too young to accommodate the recently observed old stars with age $\gtrsim 12\mathrm{Gyr}$~\citep{Vandenberg2014, GAIA_Age}. Thus, it is not very meaningful to ask whether $\Lambda$ exists, if one assumes, a priori, that $\Lambda$ is the only possibility.

The philosophy adopted in~\citet{Huang:2020mub}, and here, is to be as blind as possible when modelling the late universe. For the purpose of testing cosmic acceleration, most dark-energy type parameterizations could to some extent bias the measure, because the dark energy models are designed purposely to describe an accelerating universe. \citet{Huang:2020mub} proposed to describe the cosmic expansion history with a simple and almost model-independent Parameterization based on the cosmic Age (PAge). The PAge approximation is beyond the dark energy concept, and covers a broad class of models at the background level. Compared to the usual approach of doing model-by-model comparisons of Bayesian evidences, scanning the PAge space, as will be done in this work, is a much neater method.

The main purpose of this work is to show that, despite the aforementioned unsettled debates, cosmic acceleration is a very robust observational fact beyond supernova data, CMB data, distance-ladder measurement of $H_0$, weak lensing of galaxies, and even the concept of dark energy. This is done by navigating the PAge space with geometric information from a few other cosmological probes that will be introduced in Sec.~\ref{sec:data}.

Throughout the article we use natural units $c=\hbar=1$, a subscript ``0'' for quantities at redshift zero, and a dot to denote the derivative with respect to cosmological time $t$. For instance, $H_0$ and $t_0$ are the Hubble constant (current value of the Hubble expansion rate $H$) and the age of the universe, respectively. The deceleration parameter is defined as $q\equiv -\frac{a\ddot a}{\dot a}$, where $a$ is the scale factor.
%The Newton's constant is denoted as $G_N$ and the reduced Planck mass is defined as $M_p\equiv \frac{1}{\sqrt{8\pi G}}$.
For the rest standard symbols, such as the spatial curvature parameter $\Omega_k$ and the baryon density parameter $\Omega_b$, the reader is referred to e.g. \citet{Aghanim:2018eyx}, or any modern cosmology textbooks.

\section{PAge approximation \label{sec:page}}

For most cosmological models, the dimensionless combination $Ht$ varies slowly and smoothly. \citet{Huang:2020mub} proposed to approximate $Ht$ as a quadratic function of $t$. Assuming the matter-dominated asymptotic behavior $\left.Ht\right\vert_{t\rightarrow 0^+}\rightarrow \frac{2}{3}$, the cosmic expansion history can be approximately written as
\begin{equation}
  \frac{H}{H_0} = 1 + \frac{2}{3}\left(1-\eta\frac{H_0t}{\page}\right)\left(\frac{1}{H_0t} - \frac{1}{\page}\right), \label{eq:page}
\end{equation}
where $\page = H_0t_0$ is the age parameter, and $\eta$ measures the deviation from Einstein de-Sitter universe (flat CDM model). We only consider the physical region where $\eta<1$ \citep{Huang:2020mub}. Note that the very short period of radiation dominated epoch is ignored in PAge approximation.

The easiest way to map a physical model into PAge space is to match the deceleration parameter $q$ at some characteristic time. For simplicity and laziness (and to show the robustness of PAge), we only match $q$ at redshift zero. By taking derivative of Eq.~\eqref{eq:page}, one finds $\eta = 1- \frac{3}{2}\page^2(1+q_0)$. For flat $\Lambda$CDM model with $\Omega_m=0.3$, for instance, we have $\page = H_0t_0 = 0.964$ and $q_0=-0.55$, and hence $\eta = 0.373$.

    \begin{table*}
      \caption{\label{tab:pages} PAge approximation: maximum relative errors in angular diameter distance ($0 < z \le 2.5$)}\centering
    \begin{tabular}{lllll}
    \hline\hline

    model & parameters & $\page$ & $\eta$ & $\max\left\vert\frac{\Delta D_A}{D_A}\right\vert$\\
    \hline
    CDM &  $\Omega_m=1$& $\frac{2}{3}$ & $0$ & $0$ \\
    nonflat CDM &  $\Omega_m=0.3, \Omega_k=0.7$& $0.809$ & $-0.128$ & $0.011$ \\    
    flat $\Lambda$CDM & $\Omega_m=0.3$ & $0.964$ & $0.373$ & $0.0045$\\
    nonflat $\Lambda$CDM & $\Omega_m=0.5, \Omega_k = 0.2$ & $0.797$ & $0.0955$ & $0.0013$\\    
    flat $w$CDM & $\Omega_m=0.3, w=-1.2$  & $0.991$ & $0.647$ & $0.0060$\\
    nonflat $w$CDM & $\Omega_m=0.33, \Omega_k = -0.25, w=-0.8$  & $0.967$ & $0.269$ &$0.014$\\
    flat $w_0$-$w_a$CDM & $\Omega_m=0.3,w_0=-1.0, w_a=0.3$ & $0.953$ & $0.387$ & $0.0025$ \\
    nonflat $w_0$-$w_a$CDM & $\Omega_m=0.25, \Omega_k=0.1, w_0=-1.2, w_a=-0.2$ & $1.009$ & $0.572$ & $0.0050$ \\    
    GCG &   $\Omega_b=0.05, A=0.75, \alpha=0.1$ & $0.956$ & $0.409$ & $0.0041$ \\
    DGP &$\Omega_m=0.3$ & $0.907$ & $0.146$ & $0.0011$\\
    $R_h=ct$ & -  & $1$ & $-\frac{1}{2}$ & $0.056$\\
    \hline
    \end{tabular}
    \end{table*}

Table~\ref{tab:pages} lists a few models and their PAge approximations, characterized by the maximum fractional errors in the angular diameter distance $D_A$ in the redshift range $0<z\le 2.5$. The $w$CDM model treats dark energy as a perfect fluid with a constant equation of state $w$.  The $w_0$-$w_a$CDM model parameterizes the dark energy equation of state as a linear function of the scale factor: $w=w_0+w_a(1-a)$~\citep{Chevallier:2000qy,Linder:2002et}. The generalized Chaplygin gas (GCG) model unifies dark energy and dark matter into one fluid, whose pressure is inverse proportional to the $\alpha$-th power of the density: $p = -\frac{A}{\rho^\alpha}$~\citep{Chaplygin, Bento:2002ps}. The braneworld model of Dvali, Gabadadze, and Porrati (DGP) is a modify gravity theory that allows leakage of gravity from extra dimensions. In the simplest scenario, the DGP gravity anomaly is characterized by a scale $r_c=\frac{1}{H_0(1-\Omega_m)}$~\citep{Dvali:2000hr}. While all the above-mentioned models are well approximated by PAge, worse cases do exist. For instance, for the $R_h=ct$ linear expansion model~\citep{Melia:2011fj}, the fractional error in $D_A$ can reach $\sim 5\%$ at $z\sim 2.5$. This is because the $R_h=ct$ model is inconsistent with PAge's basic assumption of matter domination at high redshift. However, since a large fraction of the data we use in this work is at low redshift $z\lesssim 1$, PAge may still be a reasonable approximation for $R_h=ct$.

\section{Cosmological Data \label{sec:data}}

Being an approximation at the background level, PAge does not describe the growth of cosmological structures. Thus, we only use geometric information to constrain the PAge parameters. The probes we use in this work are baryon acoustic oscillations (BAO), strong gravitational lens time delay of quasars (SLTD), and passively evolving early galaxies as cosmic chronometers (CC). 

BAO as a standard ruler measuring the geometry of the late-time universe is considered to be very robust and model-independent. The BAO analysis involves a baryon acoustic scale $r_d$, which in our analysis is treated as a free parameter to reduce model dependence. We combine the geometric information from 6dF Galaxy Survey (6dFGS) and Sloan Digital Sky Survey (SDSS), as shown in Table~\ref{tab:bao}. The correlation matrices, which are needed for computation of $\chi^2$, can be found in the original publications~\citep{Beutler:2011hx, Ross:2014qpa, Alam:2016hwk, Bourboux:2017cbm, Ata:2017dya}. 

\begin{table}
    \caption{\label{tab:bao} BAO geometric constraints. Here $D_M=(1+z)D_A$ is the comoving angular diameter distance. The volume averaged scale $D_V$ is defined as $D_V\equiv\left(\frac{z D_M^2}{H}\right)^{1/3}$.}\centering
    \begin{tabular}{llll}
    \hline\hline
    redshift & measurement & value & survey\\
    \hline
    0.106 & $r_d/D_V$    & $0.336\pm 0.015$     & 6dFGS\\
    0.15  & $D_V/r_d$    & $4.466\pm 0.1681$    & SDSS DR7\\
    0.38  & $D_M/r_d$    & $10.27\pm 0.1489$    & SDSS DR12\\
    0.38  & $100 Hr_d$   & $4.018\pm 0.09366$   & SDSS DR12\\
    0.51  & $D_M/r_d$    & $13.38\pm 0.1827$    & SDSS DR12\\
    0.51  & $100Hr_d$    & $4.456\pm 0.09366$   & SDSS DR12\\
    0.61  & $D_M/r_d$    & $15.45\pm 0.2165$    & SDSS DR12\\
    0.61  & $100Hr_d$    & $4.796\pm 0.1035$    & SDSS DR12\\
    1.52  & $D_V/r_d$    & $26.01\pm 0.9948$    & SDSS DR14\\    
    2.40  & $D_M/r_d$    & $36.6\pm 1.2$        & SDSS DR12\\
    2.40  & $\frac{1}{Hr_d}$ &$8.94\pm 0.22$    & SDSS DR12\\
    \hline
    \end{tabular}
    \end{table}

In recent years, thanks to the rapidly advancing techniques of modelling time delay lens systems, SLTD has become a powerful tool to measure the angular diameter distance $D_A$ and the time-delay distance
\begin{equation}
  D_{\Delta t} = (1+z_d)\frac{D_d D_s}{D_{ds}},
\end{equation}
where $z_d$ is the redshift of lens; $D_d$ and $D_s$ are the angular diameter distance to the lens and the source, respectively; $D_{ds}$ is the angular diameter distance between the lens and the source.
We adopt five lens systems from the $H_0$ Lenses in COSMOGRAILS's Wellspring (H0LiCOW) Project and one lens system from the Strong lensing at High Angular Resolution Program (SHARP)~\citep{Wong:2019kwg}. Detailed information about the data can be found at \url{http://shsuyu.github.io/H0LiCOW/site/h0licow_data.html}.

\begin{figure}
\centering
\includegraphics[width=\columnwidth]{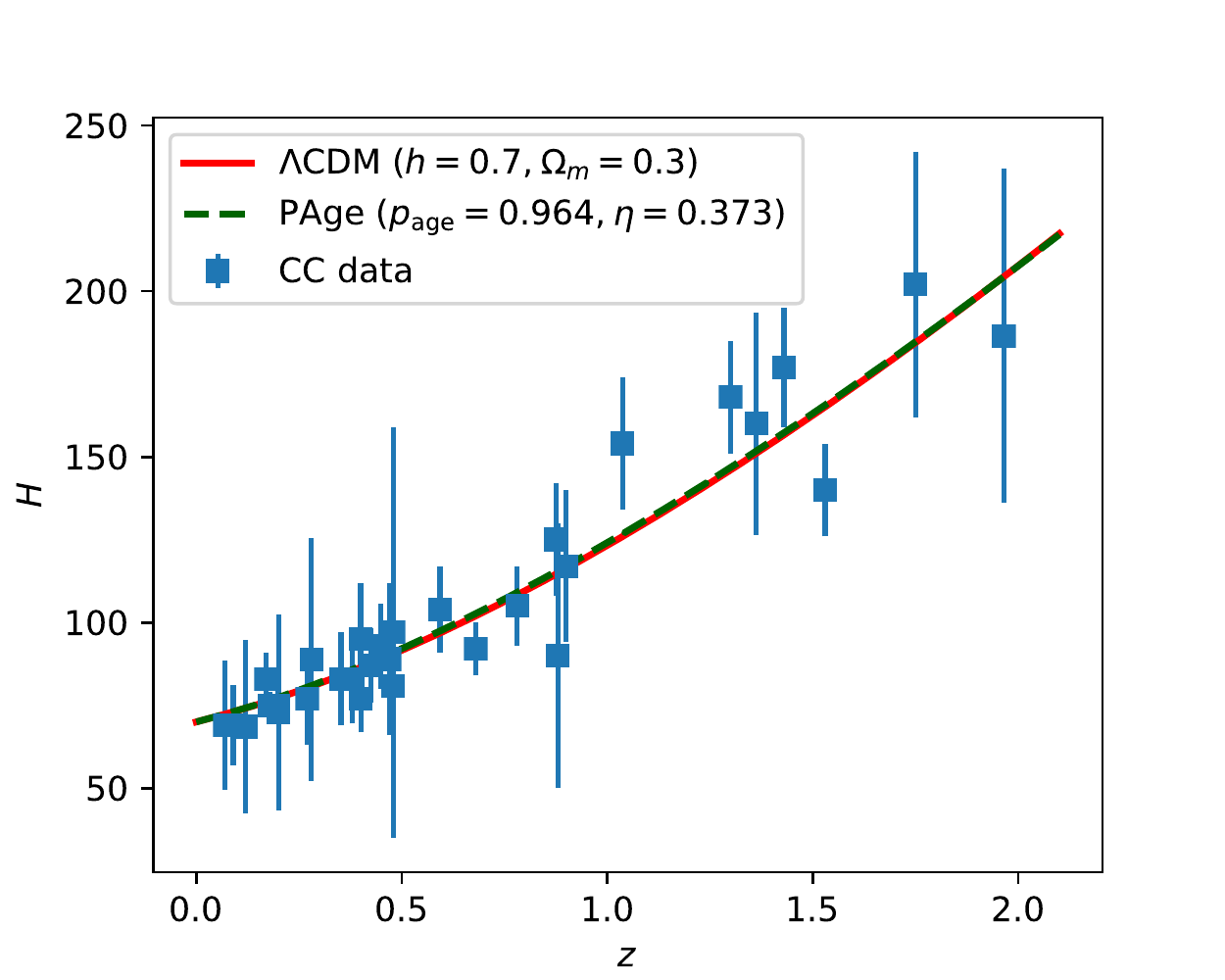}
\caption{Cosmic chronometer data. The red solid line is prediction of $\Lambda$CDM model with $\Omega_m=0.3, h=0.7$, and the green dashed line is its PAge approximation. \label{fig:Hz}} 
\end{figure}

The galaxies passively evolving on a time much longer than their age difference can be used as a cosmic chronometer~\citep{Jimenez:2001gg}. The age difference between two ensembles of old galaxies at different redshifts gives a direct estimation of the Hubble parameter $H\approx -\frac{1}{1+z}\frac{\Delta z}{\Delta t}$. We use 31 such measurements of $H(z)$~\citep{Simon:2004tf,Stern:2009ep,Zhang:2012mp,Moresco:2012jh,Moresco:2015cya,Moresco:2016mzx,Ratsimbazafy:2017vga}, which are shown in Figure~\ref{fig:Hz}. For comparison, we also plot the $H(z)$ prediction of a typical $\Lambda$CDM model with $\Omega_m=0.3, h=0.7$ and its PAge approximation.

\section{Results}

We run Monte Carlo Markov Chain (MCMC) calculations with BAO, SLTD, CC and the their combination (All), respectively. Uniform priors are applied on $h\in [0.55, 0.85]$, $\page \in [0.5, 1.5]$, $\eta \in [-2, 1]$, and $ r_dh \in [0, 1\mathrm{Gpc}]$. In addition, following ~\citet{Huang:2020mub}, we use a cosmic age bound $t_0> 12\mathrm{Gyr}$ in all cases. \modify{The chains and analysis tools are uploaded to \url{https://zenodo.org/record/4065034} to allow future researchers to reproduce our results.}

\begin{figure}
\centering
\includegraphics[width=\columnwidth]{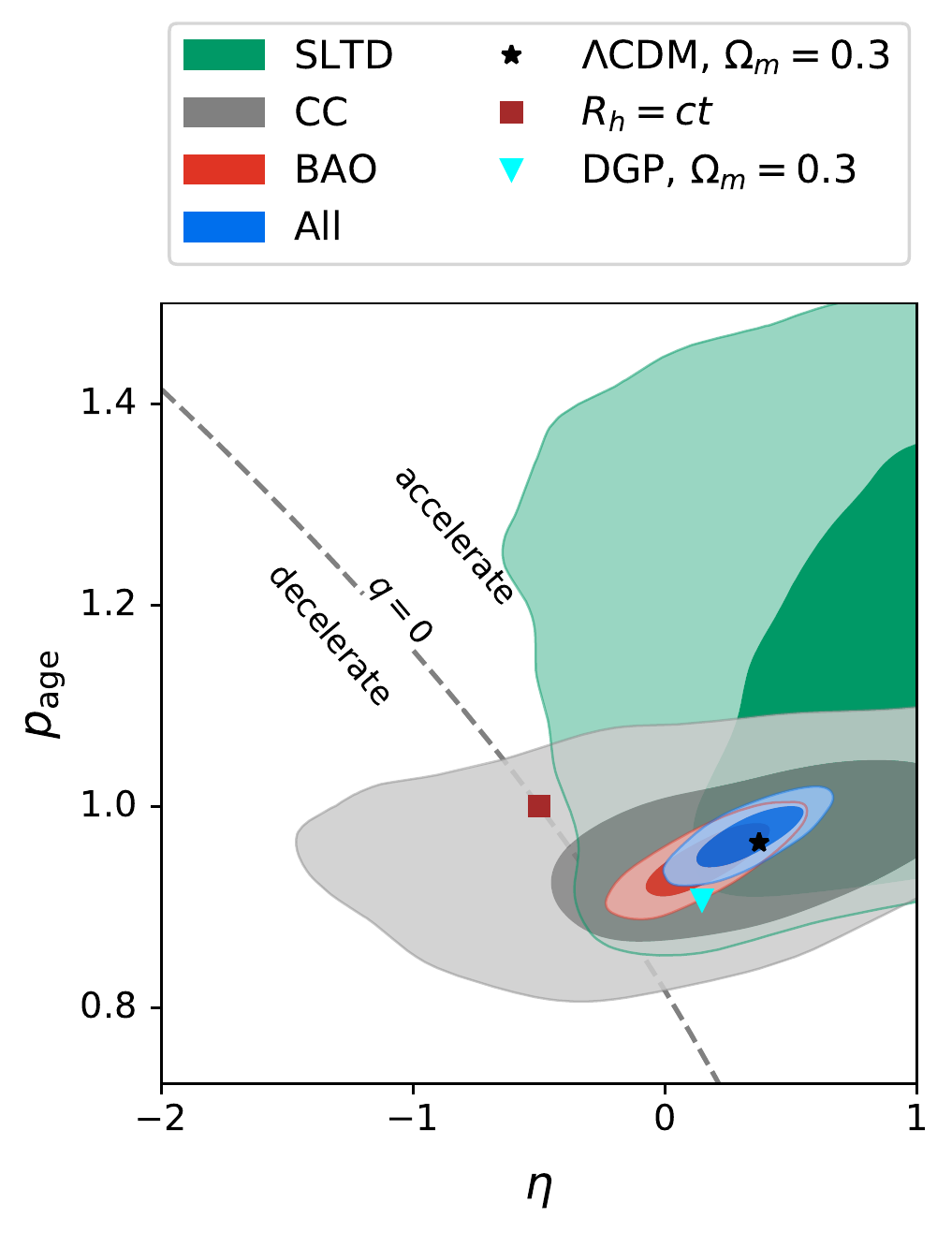}
\caption{Marginalized constraints on PAge parameters. The inner and outer contours enclose $68\%$ and $95\%$ confidence regions, respectively. \modify{The dotted violet contours show the effect of switching on the $\Omega_k$ freedom for the BAO+SLTD+CC case.} \label{fig:flat}}
\end{figure}

We firstly consider a spatially flat universe whose geometry is fully characterized by $h$, $\page$ and $\eta$. The marginalized constraints on $\page$ and $\eta$, with BAO, SLTD, CC and their combination are shown in Figure~\ref{fig:flat}. We find that CC alone does not imply cosmic acceleration, while BAO and SLTD both favor cosmic acceleration at $\sim 2.5\sigma$ level. With all the data combined together, we obtain a derived posterior of the deceleration parameter $q_0=-0.514 \pm 0.116$, a $\sim 4.4\sigma$ detection of cosmic acceleration. For the Hubble constant, we obtain $H_0=70.7\pm 1.9\, \mathrm{km\, s^{-1}Mpc^{-1}}$ that is consistent with both the local distance-ladder measurement and the CMB+$\Lambda$CDM result. The derived posterior of the baryon acoustic scale $r_d = 142.2\pm 2.7\,\mathrm{Mpc}$ is consistent with CMB + $\Lambda$CDM value $147.18\pm 0.29\,\mathrm{Mpc}$~\citep{Aghanim:2018eyx}.

Proceeding to the non-flat case with a uniform prior $\Omega_k\in [-1, 1]$, we find that the marginalized constraint on $\page$ and $\eta$ are not much affected by the addition of $\Omega_k$ freedom, \modify{as Figure~\ref{fig:flat} visually indicates.} The detection of cosmic acceleration remains at a $\sim 4.1\sigma$ level. Because the spatial curvature has almost no impact around the local universe, we can still compare the posterior of the Hubble constant $H_0=69.4\pm 1.9\hunit$ with the distance ladder measurement, and again find no significant discrepancy.

\begin{figure}
\centering
\includegraphics[width=\columnwidth]{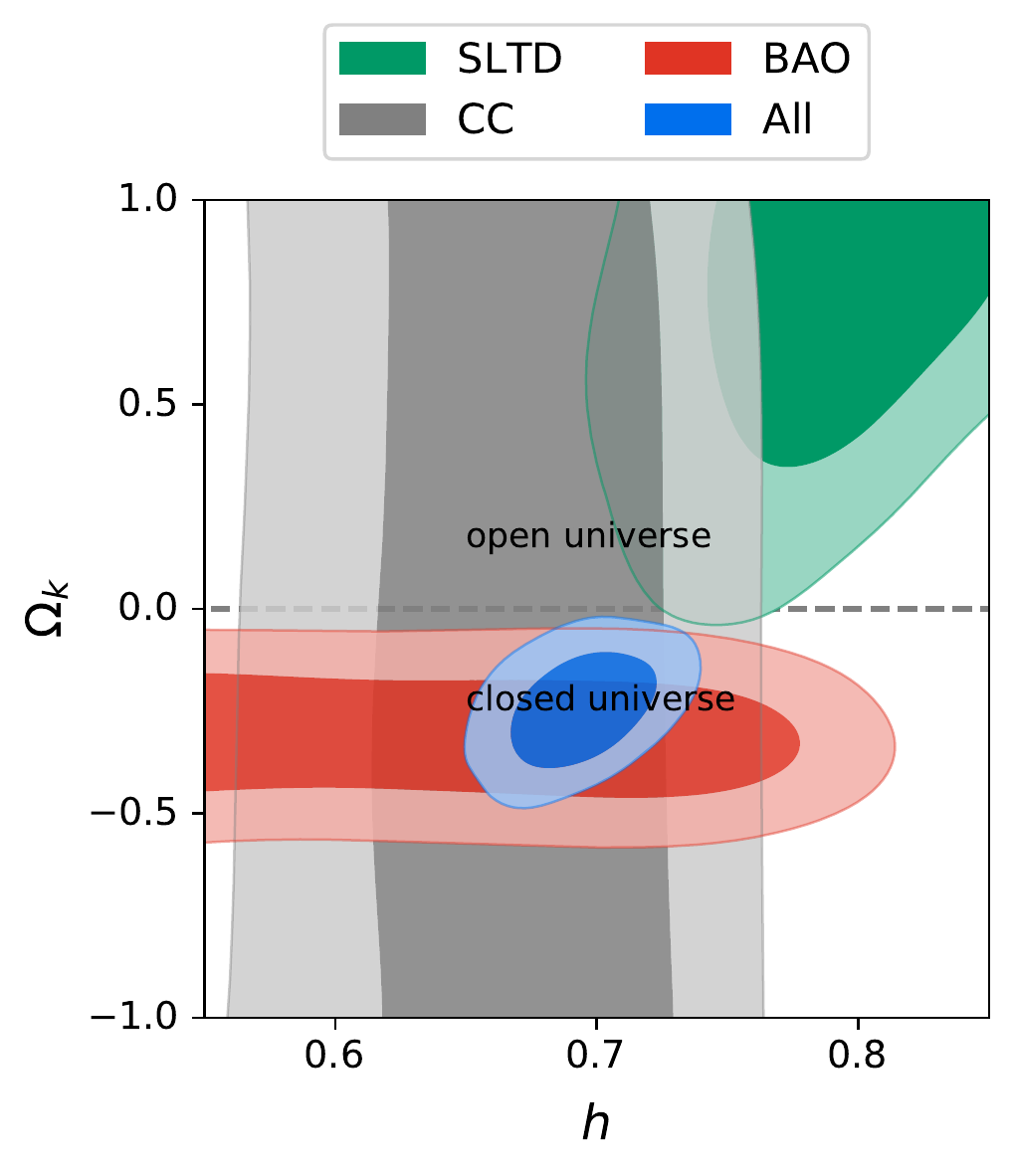}
\caption{Marginalized constraint on the reduced Hubble constant $h$ and the curvature parameter $\Omega_k$. The $\Omega_k$ tension between SLTD and BAO is about $\sim 3\sigma$.\label{fig:omk}}
\end{figure}

When $\Omega_k$ is allowed to vary, the data do not favor, however, a spatially flat universe.  Indeed, as shown in Figure~\ref{fig:omk}, SLTD favors an open universe with $\Omega_k > 0.14$ ($95.4\%$ confidence level), and BAO tends to pick up a closed universe with $\Omega_k = -0.32 \pm 0.12$. This is a novel $\gtrsim 3\sigma$ tension that does not appear in the nonflat $\Lambda$CDM framework, as we checked separately. 

If we ignore the internal tension between BAO and SLTD, a joint analysis with BAO+SLTD+CC yields $\Omega_k = -0.242\pm 0.095$, a $2.5\sigma$ preference for a closed geometry. The best-fit cosmology in this case is $(h=0.693, \Omega_k=-0.258, \page=0.972, \eta=0.281)$, which is in the proximity of the non-flat $w$CDM example in Table~\ref{tab:pages}. Going beyond $\Lambda$CDM is a key elements to see the prefernce of negative $\Omega_k$ from BAO. Indeed, hints of closed geometry, although not at a statistically significant level, have been seen for non-flat $w$CDM model in recent works that use similar data sets~\citep{Khadka:2019njj, Khadka:2020vlh}.

\section{Conclusions and Discussions \label{sec:conclu}}

Using the robust PAge approximation, we studied the cosmic expansion history with geometric information from a few cosmological probes {\it excluding} supernova, galaxy weak lensing, CMB, and distance-ladder measurement of $H_0$. We find $\gtrsim 4\sigma$ detection of cosmic acceleration for both flat and nonflat universes. The accelerated expansion of the late universe is therefore a well established and almost model-independent observational fact that can hardly be overturn by the recent debates on supernova magnitude evolution and Hubble tension.

In the nonflat universe case, we found that BAO alone prefers a closed universe at $2.7\sigma$ confidence level, but SLTD measurements favors an open universe at $\gtrsim 2\sigma$ level. The $\gtrsim 3\sigma$ tension between BAO and SLTD has not been discussed in previous works, where a CMB (or $\Lambda$CDM) prior is often assumed. However, the CMB constraint on $\Omega_k$ is model dependent, and relies on the CMB lensing reconstruction that interplays with the late universe structures. Indeed, the Planck temperature and polarization data,  if not combined with lensing reconstruction, actually favors a closed universe with $\Omega_k=-0.044^{+0.018}_{-0.015}$~\citep{Aghanim:2018eyx}. See also \citet{Handley19, DiValentino:2019qzk} for a more detailed discussion.

\begin{figure}
\centering
\includegraphics[width=\columnwidth]{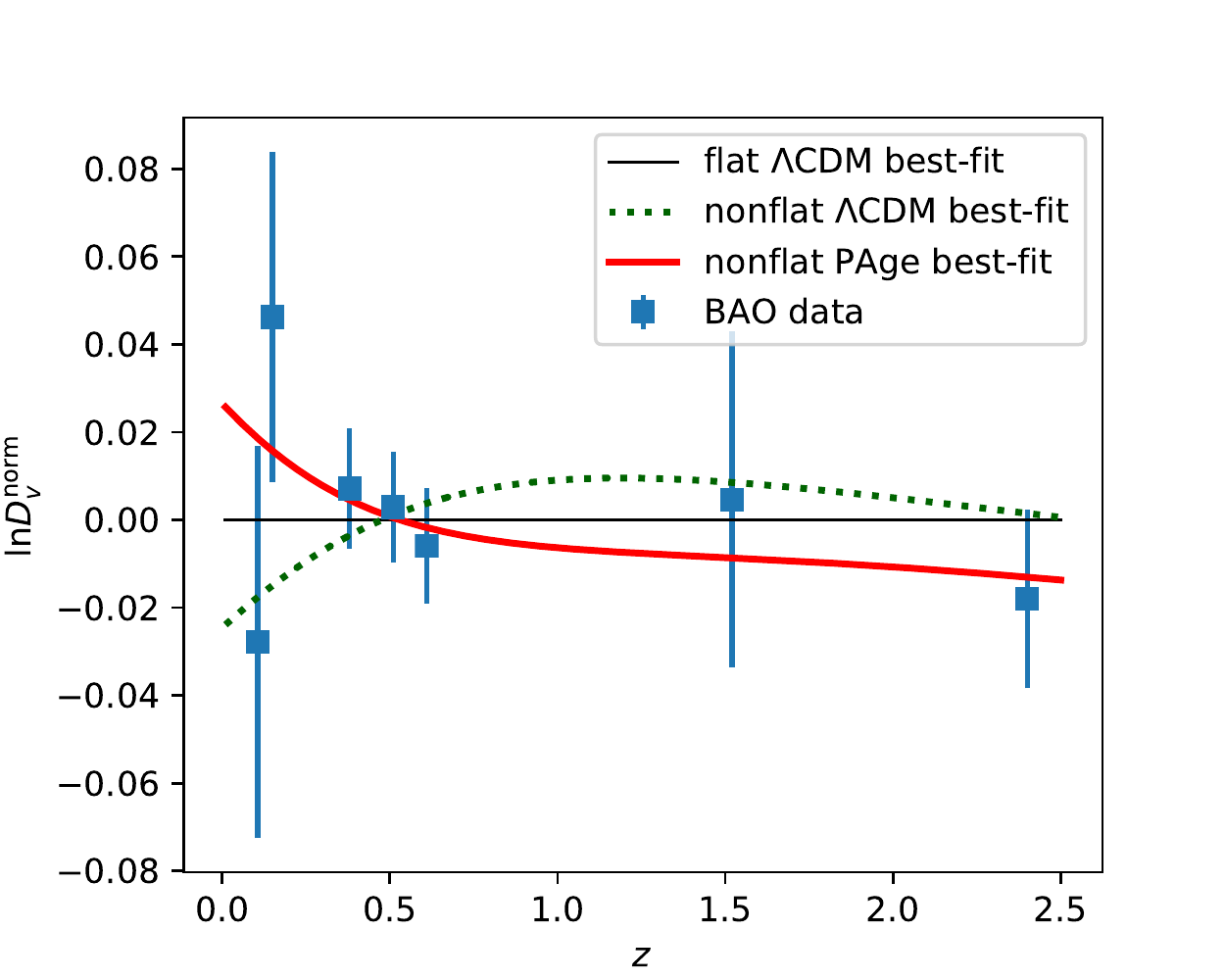}
\caption{Fitting of the converted BAO $\ln D_V$ data. For better readibility we normalized $D_V$ with the theoretical $D_V$ in the best-fit flat $\Lambda$CDM model.\label{fig:bao}}
\end{figure}

\modify{The finding that BAO alone prefers a closed PAge universe at $2.7\sigma$ is in contrast to other BAO analyses in the nonflat $\Lambda$CDM framework, where no significant detection of $\Omega_k$ has been found~\citep{Alam:2016hwk, Handley19}. There are a few differences between our analysis and the standard ones. For instance, we have avoided using the more model-dependent redshift space distortion information, i.e., the constraints on $f\sigma_8$, where $f$ is the linear growth rate. To our best understanding, however, the major driving force for a negative $\Omega_k$ is {\it not} the minor details of data selection. It is that the extra degree of freedom in nonflat PAge allows beyond-$\Lambda$CDM $D_M(z)$ and $H(z)$ trajectories that fit the data better. The improvement of fitting is sufficiently significant to beat the Occam's razor and to give a preference of nonzero $\Omega_k$. Although it is difficult to visualize the correlated data points in the high-dimensional $D_M(z)$-$H(z)$ space, we can compress the information by converting the constraints on $D_M$ and $H$ to constraints on the volume averaged $D_V\equiv\left(\frac{z D_M^2}{H}\right)^{1/3}$, with the correlation between $D_M$ and $H$ taken into account. In Figure~\ref{fig:bao} we show how the inclusion of $\Omega_k$  improves the fitting to the data, in $\Lambda$CDM and PAge respectively. In the PAge case, a better improvement of fitting leads to a stronger preference of nonzero $\Omega_k$.}

BAO peaks can be affected by the nonlinear structure of the universe, which cannot be accounted for in a truly model-independent way. The marginalization of shape information and the calculation of covariance matrix often involve simulations with a fiducial cosmology. However, empirically the marginalized distance constraints are  found to be insensitive to the fiducial cosmology~\citep{Alam:2016hwk}. Thus, unless the growth of small-scale structures is far beyond our understanding, the BAO constraints on cosmological parameters are unlikely to be significantly biased.

The complex modelling of time delay lens systems contains more potential sources of systematics. \citet{Millon_20} investigated the impact of stellar kinematics, line-of-sight effects, and deflector mass model on the inference of cosmological models. No significant biases on cosmological parameters were found. Neither does the recent detailed report on blind testing of SLTD techniques reveal any source that can significantly bias the cosmological parameters~\citep{Ding:2020jmg}. Some recent studies, however, suggest that the SLTD uncertainties increase if more flexible mass models are considered~\citep{Birrer:2020tax, Denzel:2020zuq}.

If not due to unknown systematics, the $\Omega_k$ tension could be a hint of complex expansion history that is not captured by the PAge approximation. After all, simple extensions of $\Lambda$CDM do not seem to resolve the Hubble tension, neither~\citep{Miao_2018, Guo_2019, RecombH0, WaveletH0}. Also because the $\Omega_k$ tension is a purely low-redshift phenomenon, it cannot be resolved by models that revise early universe physics, such as early dark energy~\citep{Karwal_2016,Alexander_2019,Poulin_2019}, extra relativistic species~\citep{D_Eramo_2018, Benetti_2017, Benetti_2018,Graef_2019,Carneiro_2019}, primordial magnetic field~\citep{Jedamzik_2020}, a background $T_0$ shift~\citep{Ivanov20, Bengaly20}, and variation of fundamental constants during recombination~\citep{Hart:2019dxi}. There seems to be no known viable model, even when great complexity is allowed, that can simultaneously resolves the $H_0$ tension, the $\sigma_8$ tension and the $\Omega_k$ tension. The challenge may not be trivial, as we conclude.

\section{Acknowledgments}

This work is supported by the National Natural Science Foundation of China (Grant No. 12073088) and Sun Yat-sen University Research Starting Grant 71000-18841232.

\end{document}